\documentclass[superscriptaddress,showpacs,twocolumn,preprintnumbers,amsmath,amssymb]{revtex4}
\usepackage{graphicx}
\usepackage{dcolumn}
\usepackage{bm}
\usepackage{amsmath}
\usepackage{latexsym}
\begin{document}
\title{RNA polymerase motors: dwell time distribution, velocity and dynamical phases} 
\author{Tripti Tripathi}
\affiliation{Physics Department, Indian Institute of Technology, Kanpur 208016, India.}
\author{Gunter M. Sch\"{u}tz}
\affiliation{Institut f\"{u}r Festk\"{o}rperforschung, Forschungszentrum J\"{u}lich GmbH, 52425 J\"{u}lich, Deutschland and Interdisciplinary Center for Complex Systems, University of Bonn, Germany.}
\author{Debashish Chowdhury{\footnote{Corresponding author: debch@iitk.ac.in}}}
\affiliation{Physics Department, Indian Institute of Technology, Kanpur 208016, India.}
\date{\today}%
\begin{abstract}
Polymerization of RNA from a template DNA is carried out by a molecular 
machine called RNA polymerase (RNAP). It also uses the template as a 
track on which it moves as a motor utilizing chemical energy input. 
The time it spends at each successive monomer of DNA is random; we 
derive the exact distribution of these ``dwell times'' in our model. 
The inverse of the mean dwell time satisfies a Michaelis-Menten-like 
equation and is also consistent with a general formula derived 
earlier by Fisher and Kolomeisky for molecular motors with unbranched 
mechano-chemical cycles. Often many RNAP motors move simultaneously 
on the same track.  Incorporating the steric interactions among the 
RNAPs in our model, we also plot the three-dimensional phase diagram 
of our model for RNAP traffic using an extremum current hypothesis. 
\end{abstract}
\maketitle
\section{Introduction}

RNA polymerase (RNAP) is a molecular motor \cite{schliwa}. It moves on 
a stretch of DNA, utilizing chemical energy input, while polymerizing 
a messenger RNA (mRNA) \cite{gelles98}. The sequence of 
monomeric subunits of the mRNA is dictated by the corresponding sequence 
on the template DNA. This process of template-dictated polymerization 
of RNA is usually referred to as {\it transcription} \cite{alberts}. 
It comprises three stages, namely, initiation, elongation of the mRNA 
and termination. 

We first report analytical results on the characteristic properties 
of single RNAP motors. In our approach \cite{ttdc}, each RNAP is 
represented by a hard rod while the DNA track is modelled as a 
one-dimensional lattice whose sites represent a nucleotide, the 
monomeric subunits of the DNA. The mechano-chemistry of individual 
RNAP motors is captured in this model by assigning $m$ distinct 
``chemical'' states to each RNAP and postulating the nature of the 
transitions between these states. The dwell time of an RNAP at 
successive monomers of the DNA template is a random variable; its 
distribution characterizes the stochastic nature of the movement of 
RNAP motors.  We derive the {\it exact} analytical expression for 
the dwell-time distribution  of the RNAPs in this model.

We also report results on the collective movements of the RNAPs. 
Often many RNAPs move simultaneously on the same DNA track; because 
of superficial similarities with vehicular traffic \cite{css}, we 
refer to such collective movements of RNAPs as RNAP traffic 
\cite{Klum03,polrev,liverpool,ttdc,klumpp08}. 
Our model of RNAP traffic can be regarded as an extension of the totally 
asymmetric simple exclusion process (TASEP) \cite{schuetz} for hard rods 
where each rod can exist at a location in one of its $m$ possible chemical 
states. The movement of an RNAP on its DNA track is coupled to the 
elongation of the mRNA chain that it synthesizes. Naturally, the rate of 
its forward movement depends on the availability of the monomeric subunits 
of the mRNA and the associated ``chemical'' transitions on the dominant 
pathway in its mechano-chemical cycle.  Because of the incorporation of 
the mechano-chemical cycles of individual RNAP motors, the number of rate 
constants in this model is higher than that in a TASEP for hard rods. 
Consequently, we plot the phase diagrams of our model not in a 
two-dimensionl plane (as is customary for the TASEP), but in a 3-dimensional 
space where the additional dimension corresponds to the concentration of 
the monomeric subunits of the mRNA. 

\section{Model}

\begin{figure}[ht]
\includegraphics[angle=-90,width=0.85\columnwidth]{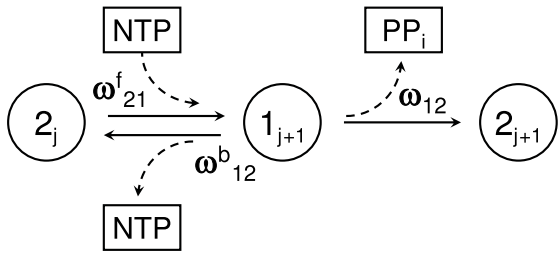}
\caption{The mechano-chemical cycle of a RNAP in our model. In the 
chemical state $1$ the RNAP is bound to pyrophosphate ($PP_i$) 
whereas no $PP_i$ is bound to it in the state $2$. The integer 
subscript $j$ labels the position of the RNAP motor on its track.}
\label{fig-model}
\end{figure}

We take the DNA template as a one dimensional lattice of length $L$ and each 
RNAP is taken as a hard rod of length $\ell$ in units of the length of a 
nucleotide. Although an RNAP covers $\ell$ nucleotides, its position 
is denoted by the {\it leftmost} nucleotide covered by it. Transcription
initiation and termination steps are taken into account by the rate 
constants $\omega_{\alpha}$ and $\omega_{\beta}$, respectively. A hard 
rod, representing an mRNA, attaches to the first site $i=1$ on the lattice 
with rate $\omega_{\alpha}$  if the first ${\ell}$ sites are not 
covered by any other RNAP at that instant of time. Similarly, an mRNA 
bound to the rightmost site $i=L$ is released from the system, with rate 
$\omega_{\beta}$. We have assumed hard core steric interaction among the 
RNAPs; therefore, no site can be simultaneously covered by more than 
one RNAP.  At every lattice site $i$, an RNAP can exist in one of
two possible chemical states: in one of these it is bound with a 
pyrophosphate (which is one of the byproducts of RNA elongation reaction 
and is denoted by the symbol $PP_i$), whereas no $PP_i$ is bound to it 
in the other chemical state (see fig.\ref{fig-model}). 
For plotting our results, we have used throughout this paper  
$\omega_{12}~ = ~31.4 ~s^{-1}$,$\omega_{12}^b~=~30.0 ~s^{-1}$ and $\omega_{21}^f~=~[NTP]~\tilde{\omega}_{21}^f ~s^{-1}$,
where $[NTP]$ is concentration of nucleotide triphosphate monomers (fuel for transcription elongation) and
$\tilde{\omega}_{21}^f~=~10^{6}~M^{-1}s^{-1}$. 
\section{Dwell time distribution}

For every RNAP, the dwell time is measured by an imaginary ``stop watch'' 
which is reset to zero whenever the RNAP reaches the chemical state $2$, 
{\it for the first time}, after arriving at a new site (say, $i+1$-th 
from the $i$-th). 

Let $P_{\mu}$ be the probability of finding a RNAP in the chemical state 
$\mu$ at time $t$. The time evolution of the probabilities $P_{\mu}$ are 
given by
\begin{eqnarray}
\frac{dP_{1}}{dt} = ~\omega_{21}^f P_{2} - ~\omega_{12} ~P_{1} 
- ~\omega^{b}_{12} ~P_{1}
\label{eq-masterp1}
\end{eqnarray}
\begin{eqnarray}
\frac{dP_{2}}{dt} = ~\omega_{12} ~P_{1} - ~\omega_{21}^f ~P_{2} + ~\omega_{12}^b ~P_{1} 
\label{eq-masterp2}
\end{eqnarray}
There is a close formal similarity between the mechano-chemical cycle 
of an RNAP in our model (see fig.\ref{fig-model}) and the 
catalytic cycle of an enzyme in the Michaelis-Menten scenario \cite{dixon}.  
The states $2$ and $1$ in the former correspond to the states $E$ 
and $ES$ in the latter where $E$ represents the free enzyme while 
$ES$ represents the enzyme-substrate complex. Following the steps of 
calculation used earlier by Kuo et al. \cite{kou05} for the kinetics  
of single-molecule enzymatic reactions, we obtain the dwell time  
distribution 
\begin{eqnarray}
f(t) = \dfrac{\omega_{21}^f~ \omega_{12}}{2A} ~ [exp\{(A-B)t\} - exp\{-(A+B)t\}] 
\label{eq-ftgen}
\end{eqnarray}
where
\begin{equation}
A = \sqrt{\frac{(\omega_{12}+\omega_{12}^b+\omega^{f}_{21})^2}{4}-\omega_{12}\omega^{f}_{21}}
\label{eq-A}
\end{equation}
\begin{equation}
B =  \frac{\omega_{12}+\omega_{12}^b+\omega^{f}_{21}}{2} 
\label{eq-B}
\end{equation} 

\begin{figure}[ht]
(a)\\
\includegraphics[width=0.85\columnwidth]{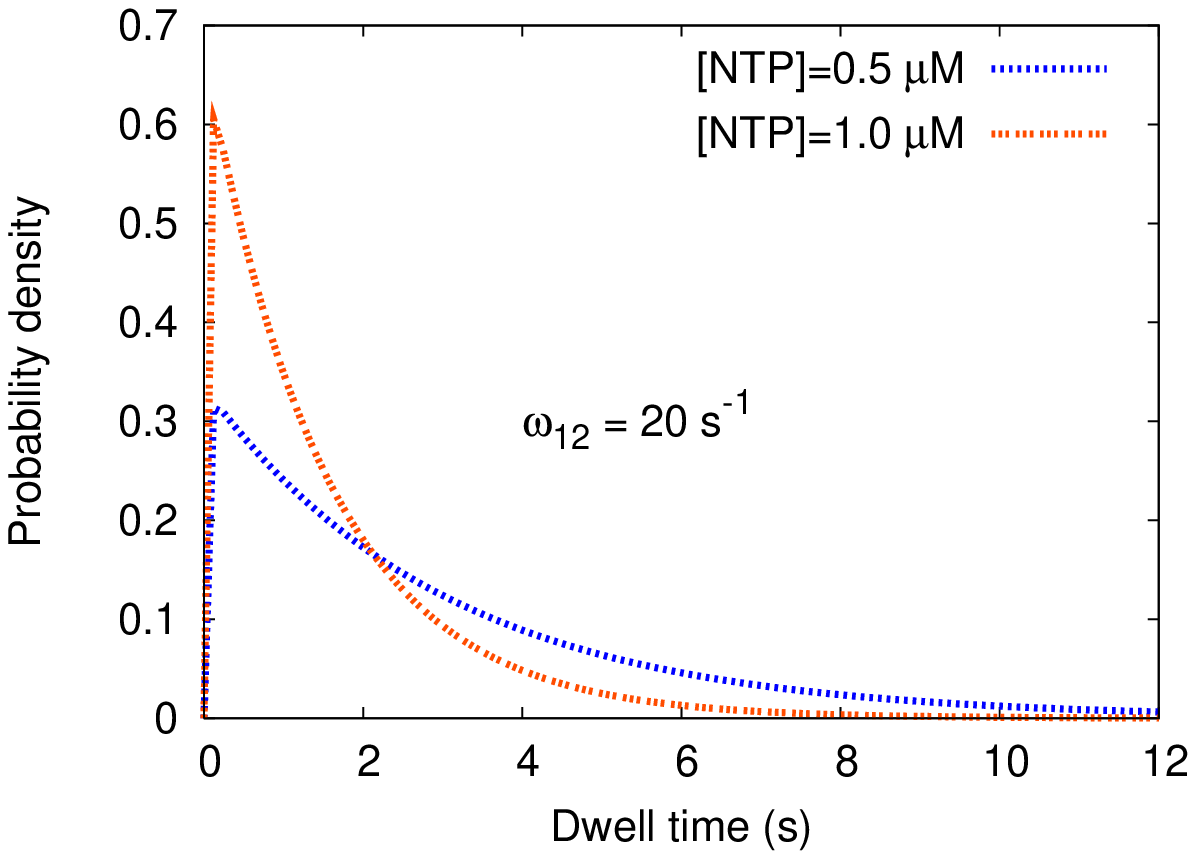}\\ 
\vspace{1cm}
(b) \\
\includegraphics[width=0.85\columnwidth]{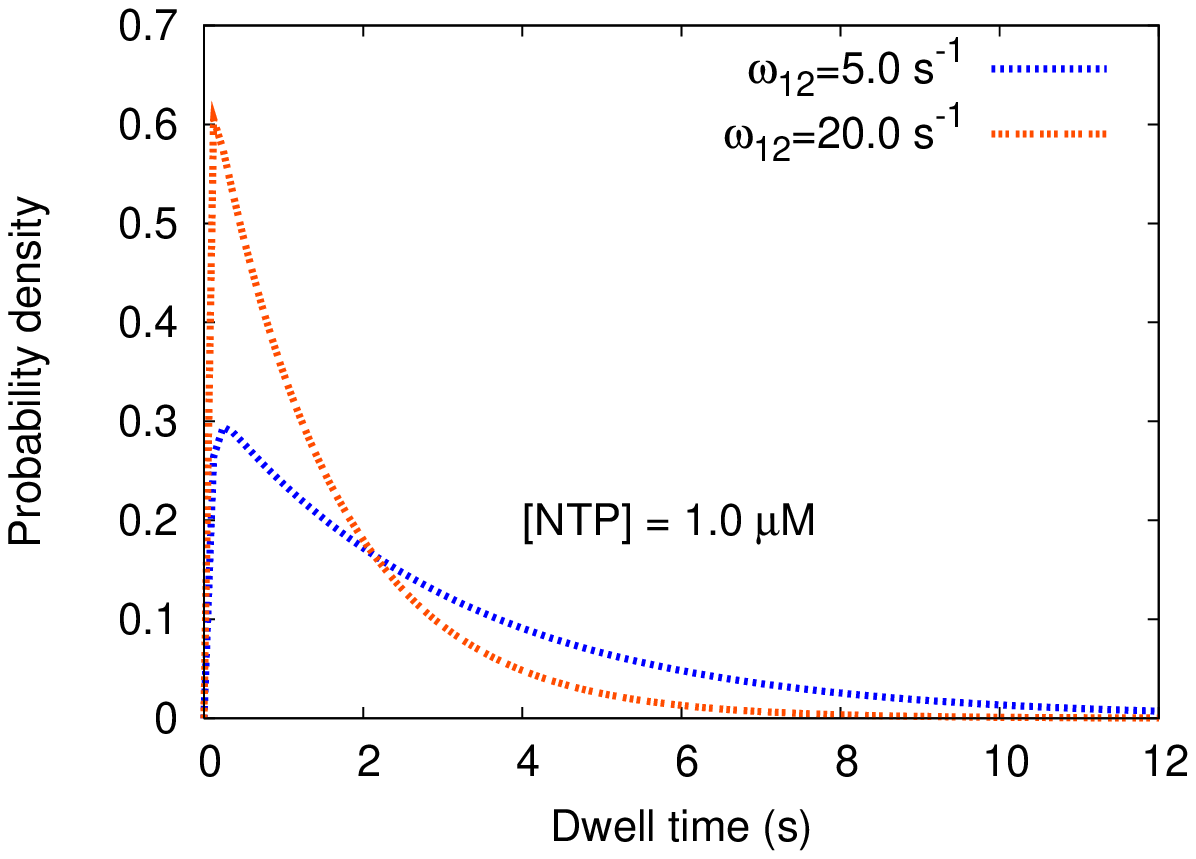}
\caption{The dwell time distribution of a single RNAP is our model 
is plotted for (a) two different values of NTP concentration, keeping 
the rate of PP$_i$ release (and, hence, $\omega_{12}$) fixed, and (b) 
two different values of $\omega_{12}$, keeping the NTP concentration 
fixed.
}
\label{fig-ft}
\end{figure}
The dwell time distribution (\ref{eq-ftgen}) is plotted in fig.\ref{fig-ft}. 
Depending on the magnitudes of the rate constants the peak of the 
distribution may appear at such a small $t$ that it may not possible to 
detect the existence of this maxium in a laboratory experiment. In 
that case, the dwell time distribution would appear to be purely a 
single exponential \cite{adelman,shund,abbon}. It is worth pointing 
out that our model does not incorporate backtracking of RNAP motors 
which have been observed in the {\it in-vitro} experiments 
\cite{shaevitz,galburt}. It has been argued by some groups \cite{neuman} 
that short transcriptional pausing is distinct from the long pauses 
which arise from backtracking. In contrast, some other groups \cite{depken} 
claim that polymerase backtracking can account for both the short and 
long pauses. Thus, the the role of backtracking in the pause distribution 
remain controversial. Moreover, it has been demonstrated that a polymerase 
stalled by backtracking can be re-activated by the ``push'' of another  
closely following it from behind \cite{nudler03a}. Therefore, in 
the crowded molecular environment of intracellular space, the occurrence 
of backtracking may be far less frequent that those observed under 
{\it in-vitro} conditions. Our model, which does not allow backtracking, 
predicts a dwell time distribution which is qualitatively very similar to 
that of the short pauses provided the most probable dwell time is shorter 
than 1 s.

From equation (\ref{eq-ftgen}) we get  the inverse mean dwell time
\begin{equation}
\dfrac{1}{\langle t \rangle} = \dfrac{\tilde{V}_{max}}{1+\dfrac{\tilde{K}_M}{[NTP]}}
\label{eq-avtgen} 
\end{equation}
where $\tilde{V}_{max} = \omega_{12}$ and 
${\tilde{K}_M} = (\omega_{12}+\omega_{12}^b)/\tilde{\omega}^f_{21}$. 
The form of the expression (\ref{eq-avtgen}) is identical to the Michaelis-Menten 
formula for the average rate of an enzymatic reaction. It describes
the slowing down of the ``bare''  elongation progress of an RNAP due to the NTP reaction cycle
that it has to undergo. The unit of velocity is $nucleotide/second$.

The fluctuations of the dwell time can be computed from the second moment
\begin{eqnarray}
 \langle t^2 \rangle &=& \int_{0}^{\infty} ~t^2 ~f(t)~ dt\\
&=&~\dfrac{2~\left[ \left(\omega_{12}+\omega_{12}^b+\omega_{21}^f \right)^2 ~-~\omega_{12}~\omega_{21}^f \right] }{\left(\omega_{12}~\omega_{21}^f  \right)^2 }
\end{eqnarray}
of the dwell time distribution. We find the randomness parameter 
\cite{kou05,block,ttthesis}
\begin{eqnarray}
 \gamma & = & \dfrac{\langle t^2 \rangle~-~\langle t \rangle^2}{\langle t \rangle^2}\\
 & = & \dfrac{\left(\omega_{12}+\omega_{12}^b+\omega_{21}^f \right)^2 ~-~2~\omega_{12}~\omega_{21}^f}{\left(\omega_{12}+\omega_{12}^b+\omega_{21}^f \right)^2}
\label{eq-ranpar}
\end{eqnarray}

\begin{figure}[ht]
\centering
\includegraphics[width=0.85\columnwidth]{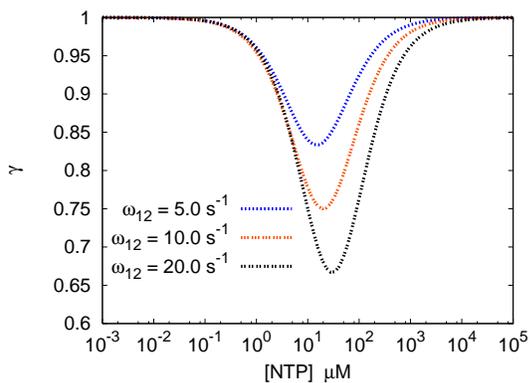}\\[0.5cm]
\caption{The randomness parameter $\gamma$ is plotted against NTP 
concentration for three values of the parameter $\omega_{12}$.
}
 \label{fig-ranpar}
\end{figure}

Note that, for a one-step Poisson process $f(t) = \omega exp(-\omega t)$, 
$\gamma = 1$.
The randomness parameter $\gamma$, given by (\ref{eq-ranpar}), is plotted 
against the NTP concentration in fig.\ref{fig-ranpar} for three different 
values of $\omega_{12}$. At sufficiently low NTP concentration, $\gamma$ 
is unity because NTP binding with the RNAP is the rate-limiting step. As 
NTP concentration increases, $\gamma$ exhibits a nonmonotonic variation.
At sufficiently high NTP concentration, PP$_{i}$-release (which occurs 
with the rate $\omega_{12}$) is the rate-limiting step and, therefore, 
$\gamma$ is unity also in this limit. This interpretation is consistent 
with the fact that the smaller is the magnitude of $\omega_{12}$, the 
quicker is the crossover to the value $\gamma = 1$ as the NTP concentration 
is increased.

The randomness parameter yields the diffusion coefficient \cite{block}
\begin{eqnarray}
&&D~=\dfrac{\gamma}{2  \langle t \rangle }\\
&&=\dfrac{\omega_{12}\omega_{21}^f}{\omega_{12}+\omega_{12}^b+\omega_{21}^f}\left[\dfrac{\left(\omega_{12}+\omega_{12}^b+\omega_{21}^f \right)^2 -2\omega_{12}\omega_{21}^f}{2\left(\omega_{12}+\omega_{12}^b+\omega_{21}^f \right)^2} \right]\nonumber\\
\label{eqn-diffusion}
\end{eqnarray}
The expression (\ref{eqn-diffusion}) is in agreement with the general expression 
for the effective diffusion constant of a molecular motor
with unbranched mechano-chemical cycle which was first reported by Fisher and Kolomeisky \cite{fishkolo}.

\section{Phase Diagrams}
Now we will take into account the hard core steric interaction among the RNAPs which 
are simultaneously moving on the same DNA track. Equations 
(\ref{eq-masterp1} and \ref{eq-masterp2}) will be modified to
\begin{eqnarray}
\frac{dP_{1}}{dt} = ~\omega_{21}^f P_{2}~Q - ~\omega_{12} ~P_{1} 
- ~\omega^{b}_{12} ~P_{1}~Q
\label{eq-masterp12}
\end{eqnarray}
\begin{eqnarray}
\frac{dP_{2}}{dt} = ~\omega_{12} ~P_{1} - ~\omega_{21}^f ~P_{2}~Q + ~\omega_{12}^b ~P_{1}~ Q
\label{eq-masterp22}
\end{eqnarray}
Where $Q$ is conditional probability \cite{ttdc} of finding site $i+\ell$
($i-\ell-1$ for backward motion) vacant, given there is a particle at site $i$.

Due to the steric interactions between RNAP's their stationary flux $J$ 
(and hence the transcription rate) is no longer limited solely by the
initiation and release at the terminal sites of the template DNA.
We calculate the resulting phase diagram utilizing the extremum current 
hypothesis (ECH) \cite{krug91,popkov99}. 
The ECH relates the flux in the system under open boundary conditions 
(OBC) to that under periodic boundary conditions (PBC) with the same 
bulk dynamics. In this approach, one imagines that initiation and 
termination sites are connected to two 
separate reservoirs where the number densities of particles are 
$\rho_{-}$ and $\rho_{+}$ respectively, and where the particles 
follow the same dynamics as in the bulk of the real physical system.  Then
\begin{eqnarray}\label{ECH}
J  = \left\{ \begin{array}{ll}
         \mbox{max}~~ j(\rho) & \mbox{if}~~ \rho_{-}~>~\rho~>~\rho_{+}\\
         \mbox{min} ~~j(\rho)& \mbox{if}~~ \rho_{-}~<~\rho~<~\rho_{+}.\end{array} \right. \nonumber\\
\end{eqnarray}
The actual rates $\omega_{\alpha}$ and $\omega_{\beta}$ of initiation 
and termination of mRNA polymerization are incorporated by appropriate 
choice of $\rho_{-}$ and $\rho_{+}$ respectively. 

\begin{figure}[ht]
\centering
\includegraphics[width=0.85\columnwidth]{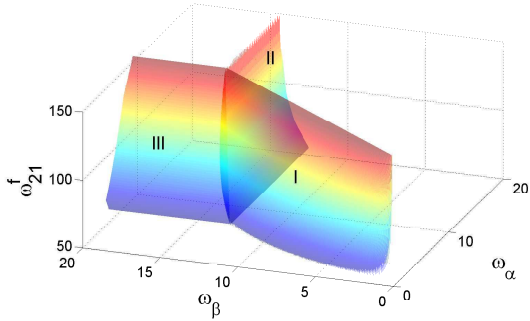}\\[0.5cm]
\caption{The 3-d phase diagram of our model for RNAP traffic. The LD 
and HD phases coexist on the surface $I$. The surfaces $II$ and 
$III$ separate the MC phase from the HD and LD phases, respectively. 
}
 \label{fig:3d}
\end{figure}
\begin{figure}[ht]
 \centering
 \includegraphics[width=0.85\columnwidth]{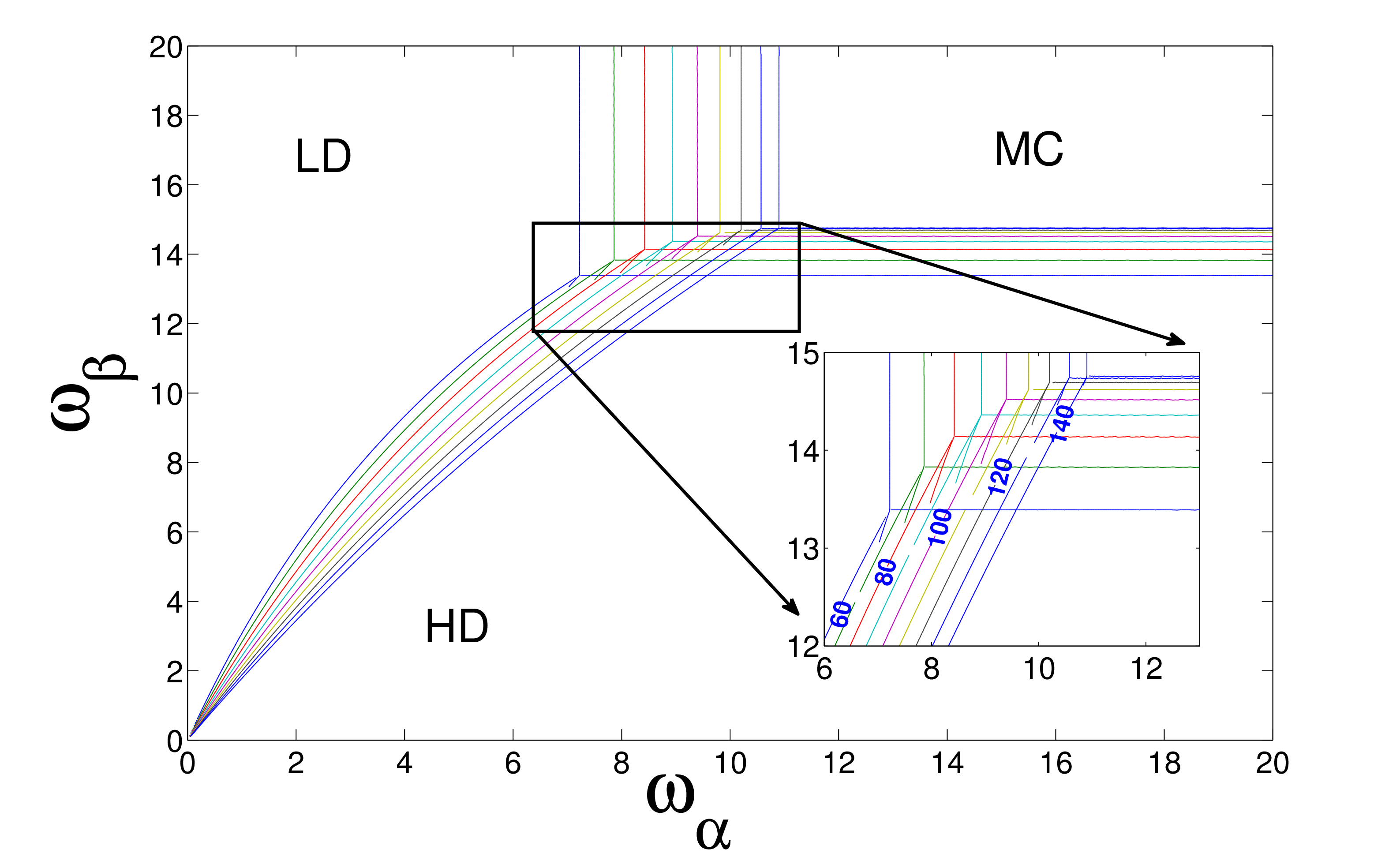}
\caption{Projections of several 2-d phase diagram of RNAP traffic on 
the $\omega_{\alpha}$-$\omega_{\beta}$ plane for several values of 
$\omega_{21}^f$. The numbers on the phase boundary lines represent 
the value of $\omega_{21}^f$. The inclined lines have LD and HD 
above and below, respectively, while the MC phase lies in the 
upper right corner. }
 \label{fig:xyplane}
\end{figure}
 
An expression for $J(\rho)$ was reported by us in Ref.~\cite{ttdc}. In 
the special case where the dominant pathway is that shown in  
Fig.~\ref{fig-model} ($\omega_{12}^b$~=~0 for further calculation as $\omega_{12}~\gg ~\omega_{12}^b$), we have  
\begin{equation}
 J(\rho)=\dfrac{\omega_{12}~\omega_{21}^f~\rho~(1-\rho~\ell)}{\omega_{12}~(1-\rho~(\ell-1))~+~\omega_{21}^f~(1-\rho~\ell)}
\label{eq:effj}
\end{equation}
The number density $\rho^*$ that corresponds to the maximum flux is 
given by the expression
\begin{equation}
\rho^*=\sqrt{\dfrac{\omega_{12}~+~\omega_{21}^f}{\ell~\omega_{12}}}\left[\sqrt{\ell \left(\dfrac{\omega_{12}~+~\omega_{21}^f}{\omega_{12}} \right) }+1 \right]^{-1}
\end{equation}
By comparing (\ref{eq:effj}) with the exact current-density relation 
of the usual TASEP for extended particles of size $\ell$ 
\cite{macdonald68,shaw03,Scho04}, which have no internal states
(formally obtained by taking the limit $\omega_{12}\to \infty$ in the 
present model), we predict that the stationary current (i.e. the 
collective average rate of translation) is reduced by the occurrence 
of the intermediate state 1 through which the RNAPs have to pass.

From (\ref{ECH}) one expects three phases, viz. a maximal-current(MC) phase with with bulk density $\rho^*$,  a
low-density phase (LD) with bulk density $\rho_{-}$,  and a high-density phase (HD) with bulk density $\rho_{+}$.
Using arguments similar to those used in Ref.~\cite{basuchow} in a similar  
context, we get \cite{ttthesis}
\begin{equation}
 \rho_{-} = \dfrac{\omega_{\alpha}~\left(\omega_{12}+\omega_{21}^f\right)}{\omega_{12}~\omega_{21}^f~+~\omega_{\alpha} \left(\omega_{12}+\omega_{21}^f\right) \left(\ell -1 \right)}
\label{eq-rhom}
\end{equation}
and 
\begin{equation}
\rho_{+}=\dfrac{\omega_{12}~\omega_{21}^f~-~\omega_{\beta}~\left(\omega_{12}+\omega_{21}^f\right)}{\omega_{12}~\omega_{21}^f~\ell~-~\omega_{\beta}~\left( \omega_{12}+\omega_{21}^f\right)(\ell~-~1)}.
\label{eq-rhop}
\end{equation}

\begin{figure}[ht]
 \centering
\includegraphics[width=0.85\columnwidth]{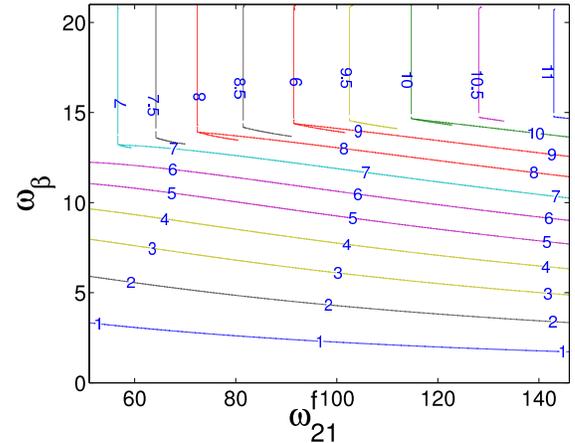}
\caption{Same as in fig.\ref{fig:xyplane} except that the projections are 
on $\omega_{21}^f$-$\omega_{\beta}$ plane for several values of 
$\omega_{\alpha}$. The inclined lines have LD and HD above and below, 
respectively. Each vertical line separates the LD phase on the left 
from the MC phase on its right.}
\label{fig:yzplane}
\end{figure}

The condition for the coexistence of the high density (HD) and low 
density (LD) phases is
\begin{equation}
 J(\rho_{-}) = J(\rho_{+}) 
\label{eq-ldhd} 
\end{equation} 
with $\rho_{-} < \rho^\ast < \rho_{+}$.
Using the expression (\ref{eq:effj}) for $J$ in (\ref{eq-ldhd}) we get
\begin{eqnarray}
\rho_+&=&\dfrac{\left(\omega_{12}~+~\omega_{21}^f\right)\left(1-\rho_-\ell \right)}{\ell \left(\omega_{12}~+~\omega_{21}^f\right)\left(1-\rho_-\ell \right)~+~\rho_- \ell \omega_{12}} .
\label{eq-rhomp}
\end{eqnarray} 
Substituting (\ref{eq-rhom}) and (\ref{eq-rhop}) into (\ref{eq-rhomp}), 
we get the equation for the plane of coexistence of LD and HD to be 
$\omega_{\beta} = f(\omega_{\alpha},\omega_{21}^{f})$ where 
\begin{eqnarray}
f(\omega_{\alpha},\omega_{21}^{f})= \dfrac{\left(\omega_{12} \omega_{21}^f \right) \omega_{12}\ell \omega_{\alpha}}{\left(\omega_{12}+\omega_{21}^f \right) \left[ \omega_{12} \omega_{21}^f-\omega_{\alpha}\left(\omega_{21}^f+\omega_{12}-\omega_{12}\ell \right) \right] }.
\label{eq:phasec}
\end{eqnarray}
In order to compare our result with the 2-d phase diagram of the TASEP 
in the $\omega_{\alpha}-\omega_{\beta}$-plane, we project 2-d cross 
sections of the 3-d phase diagram, for several different values of 
$\omega_{21}^{f}$ onto the $\omega_{\alpha}-\omega_{\beta}$-plane. 
The lines of coexistence of the LD and HD phases on this projected 
two-dimensional plane are curved , a similar curvature is also
reported by Antal and Sch\"{u}tz \cite{antal}. This is in contrast to the 
straight coexistence line for LD and HD phases of TASEP.

The bulk density of the system is guided by following equations: 
\begin{widetext}
\begin{eqnarray}
\rho(\omega_{\alpha},\omega_{\beta})  = \left\{ \begin{array}{lll}
          \rho_{-} &~ \mbox{if}~\omega_{\beta}>    
f(\omega_{\alpha},\omega_{21}^f) {\rm and} ~\omega_{\alpha} < \biggl[\dfrac{\rho_*}{1-\rho_*(\ell-1)}\biggr] \biggl[\dfrac{\omega_{12}\omega_{21}^f}{\omega_{12}+\omega_{21}^f}\biggr]~~\mbox{Low density}\\
\rho_{+} &~ \mbox{if}~\omega_{\beta}< f(\omega_{\alpha},\omega_{21}^f)   
{\rm and} ~\omega_{\beta} < \biggl[\dfrac{1-\rho_*\ell}{1-\rho_*(\ell-1)}\biggr] \biggl[\dfrac{\omega_{12}\omega_{21}^f}{\omega_{12}+\omega_{21}^f}\biggr]~~\mbox{High density } \\
\rho_{*} &~\mbox{if}~\omega_{\beta}>\biggl[\dfrac{1-\rho_*\ell}{1-\rho_*(\ell-1)}\biggr] \biggl[\dfrac{\omega_{12}\omega_{21}^f}{\omega_{12}+\omega_{21}^f}\biggr] {\rm and} ~\omega_{\alpha}> \biggl[\dfrac{\rho_*}{1-\rho_*(\ell-1)}\biggr] \biggl[\dfrac{\omega_{12}\omega_{21}^f}{\omega_{12}+\omega_{21}^f}\biggr]~~\mbox{Maximal current}.\end{array} \right. \nonumber\\
\end{eqnarray}
\end{widetext}
In Fig.~\ref{fig:3d}, we plot the 3d phase diagram. 

\begin{figure}[ht]
 \centering
\includegraphics[width=0.85\columnwidth]{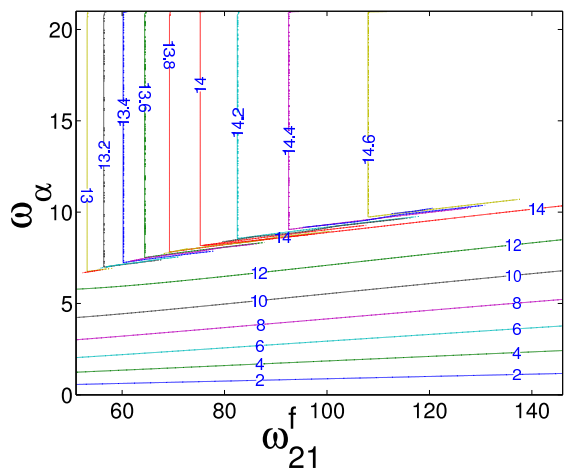}
\caption{Same as in fig.\ref{fig:xyplane} except that the projections are
on $\omega_{21}^f$-$\omega_{\alpha}$ plane for several values of 
$\omega_{\beta}$. Here the inclined lines have HD and LD, respectively, 
above and below. Each vertical line separates the HD phase on the left 
from the MC phase on its right.}
\label{fig:xzplane}
\end{figure}

In general, a plane $\omega^{f}_{21}$=constant intersects the surfaces 
I, II and III thereby generating the phase transition lines between the 
LD, HD and MC phases in the $\omega_{\alpha}$-$\omega_{\beta}$ plane. 
We have projected several of these 2d phase diagrams, each for one 
constant value of $\omega^{f}_{21}$ in figure \ref{fig:xyplane}. 
In the inset, we have shown the value of $\omega_{21}^f$ for different 
lines. We have also projected several 2d phase diagrams in the 
$ \omega_{21}^f$-$\omega_{\beta}$ plane and $\omega_{21}^f$-$\omega_{\beta}$ 
plane, respectively, in figures \ref{fig:yzplane} and \ref{fig:xzplane}. 

\section{Summary and Conclusion}

In this paper we have reported the exact dwell time distribution for a 
simple 2-state model of RNAP motors. From this distribution we have 
also computed the average velocity and the fluctuations of position and dwell time 
of RNAP's on the DNA nucleotides. These expressions are consistent with a general 
formula derived earlier by Fisher and Kolomeisky for a generic model 
of molecular motors with unbranched mechano-chemical cycles. 

Taking into account the presence of steric interactions between
different RNAP moving along the same DNA template we have plotted 
the full 3d phase diagram of a model for multiple RNAP traffic. 
This model is a biologically motivated extension 
of the TASEP, the novel feature being the incorporation of the mechano-chemical
cycle of the RNAP into the dynamics of the transcription process.
This leads to a hopping process with a dwell time distribution that is
not a simple exponential. 
Nevertheless, the phase diagram is demonstrated to follow
the extremal-current hypothesis \cite{popkov99} for driven diffusive systems.
Using mean field theory we have computed the effective
boundary densities that enter the ECH from the reaction constants
of our model. We observe that the collective average rate of translation as given by the 
stationary RNAP current (\ref{eq:effj}) is reduced by the need of the RNAP 
to go through the pyrophosphate bound state. This is a prediction that is
open to experimental test.

The 2d cross sections of this 
phase diagram have been compared and contrasted with the phase diagram 
for the TASEP.  Unlike in the TASEP, the
coexistence line between low- and high-density phase is curved for all parameter
values.
This is a signature of broken
particle-vacancy symmetry of the RNAP dynamics. The presence of this
coexistence line suggests the occurrence of RNAP ``traffic jams'' that our model
predicts to appear when stationary initiation and release of RNAP at the terminal sites of the
DNA track are able to balance each other. This traffic jam would perform an
unbiased random motion, as argued earlier on general theoretical grounds
in the context of protein synthesis by ribosomes from mRNA templates
\cite{Schu97}.\\

\noindent{\bf Acknowledgments}: This work is supported by a grant from CSIR (India).
GMS thanks IIT Kanpur for kind hospitality and DFG for partial financial support.


\end{document}